\documentclass{jps-cp}
\usepackage{txfonts} 
\usepackage{bm}
\usepackage{here}
\usepackage{mathtools}
\usepackage{color}
\usepackage{ulem}

\title{High-harmonic generation in the Rice-Mele model: Role of intraband current originating from interband transition}

\author{Kohei  \textsc{Nagai}$^{1}$, Yuta \textsc{Murakami}$^{1,2}$ and Akihisa \textsc{Koga}$^{1}$}

\inst{$^{1}$ Department of Physics, Tokyo Institute of Technology, Meguro, Tokyo 152-8551, Japan \\
$^{2}$Center for Emergent Matter Science, RIKEN, Wako, Saitama 351-0198, Japan}

\email{nagai.k@stat.phys.titech.ac.jp}

\abst{
  We consider high-harmonic generation (HHG) in the Rice-Mele model
  to study the role of the intraband current originating from
  the change of the intraband dipole via interband transition.
  This contribution, which has been often neglected in previous works,
  is necessary for the consistent theoretical formulation
  of the light-matter coupling. 
  We demonstrate that the contribution becomes crucial
  when the gap is smaller than or comparable to the excitation frequency and
  the system is close to the half filling. }

\kword{High-harmonic generation}

\begin{document}
\maketitle

\section{Introduction}
High-harmonic generation is a fundamental nonlinear optical phenomenon originating from strong light-matter coupling. It was originally observed and studied in atomic and molecular gases~\cite{Ferray,Corkum}. 
Recently, HHG was observed in solids, such as semiconductors and semimetals, which extended the scope of the HHG research to solids ~\cite{Ghimire2010Nature,Luu2015Nature,Hohenleutner2015Nature,Yoshikawa2017Science}. 
One intriguing aspect of solids is that their properties can be controlled using active parameters such as temperatures, doping and pressure~\cite{Nishidome2020Nano,Tamaya2021PRB,Uchida2022PRL,Ma2022PRA,Murakami2022a}. 
In order to correctly predict the dependence of HHG on these parameters, the consistent theoretical treatment of the light-matter coupling is necessary.
A major approach to study HHG in solids is
the semiconductor Bloch equations (SBEs) focusing on the several bands around the Fermi level~\cite{Vampa2015PRB,Luu2016PRB,Charcon2020PRB}.
However, the expression of SBEs depends on the gauges for the light and bases for electron states, which may lead to the inconsistency among the results
obtained in terms of distinct choices.
Recently, the relation between the different representations has been investigated
in detail~\cite{Wilhelm2021PRB,Yue2022Tutorial,Murakami2022b}.
There is a term in the interaband current that represents the change of the intraband dipole via interband transition
although it is often neglected in the HHG analysis
based on the well-used SBEs.
Then a question arises; in which conditions this contribution
is crucial for HHG?
To answer this question, we numerically study HHG
in the one-dimensional Rice-Mele model~\cite{Rice1982PRB}.
We discuss how important the often-neglected term is
in the system when the gap-size and doping level is systematically changed.

\section{Model and Method}
We start with the one-dimensional Rice-Mele model~\cite{Rice1982PRB} in the length gauge, whose Hamiltonian is
\begin{equation}
\hat H_{0}=\sum_{i}\frac{ Q_{x}-Q_{y}(-1)^i}{2}(c_{i}^{\dagger} c_{i+1}+h.c.)+\sum_i Q_{on}(-1)^{i}c_{i}^{\dagger}c_{i} -qE(t) \sum_i r_i c_{i}^{\dagger}c_{i}, \label{eq:H_basic}
\end{equation}
where $c_{i}^{\dagger}$ creates an electron at  the $i$th site,
$Q_{x}$ is the averaged hopping between the nearest neighboring sites,
$Q_{y}$ is the hopping alternation, and
$Q_{on}$  is the staggered onsite energy.
$q$ is the charge, $E(t)$ is the electric field, and
$r_i$ is the position of the $i$th site.
In this model, we have introduced the light-matter coupling,
assuming that in the length gauge,
the dipole matrix elements between states on different sites
are zero~\cite{Murakami2022b}.
It is known that even harmonics in the HHG spectrum disappear
when the system has the inversion symmetry.
In the case with $Q_{y}\neq 0$ and $Q_{on}\neq 0$
in the model (\ref{eq:H_basic}),
the system is not invariant under the inversion,
leading to not only odd harmonics but also even harmonics in the HHG spectrum.

In the following, we introduce three seemingly-different but essentially-equivalent representations,
which are obtained from Eq.~\eqref{eq:H_basic} by unitary transformations.

\subsection{Representation I: Dipole gauge expressed with the localized Wannier basis}
 In the dipole gauge, the light-matter coupling is taken into account through the Peierls phase as 
\begin{equation}
\hat{H}^D(t)=\sum_{i}\left(\frac{ Q_{x}-Q_{y}(-1)^i}{2}(e^{{-iqaA(t)}}c_{i}^{\dagger} c_{i+1}+h.c.)+Q_{on}(-1)^{i}c_{i}^{\dagger}c_{i}\right), \label{eq:H_dipole}
\end{equation}
where $a$ is the bond length,
and $A(t)$ is the vector potential.
Namely, $E(t)=-\partial_{t} A (t)$.
Using the Fourier transformations, the creation operators
in the $\alpha(=A, B)$ sublattice are given by
$c_{k\alpha}^{\dagger}=\frac{1}{\sqrt{N}}\sum_{i \in \alpha}e^{ikr_{i}}c_{i}^{\dagger}$.
 The Hamiltonian in the $k$-space is given as 
\begin{equation} 
\hat{H}^D(t)=\sum_{k}
\begin{pmatrix}
c_{kA}^{\dagger}~~c_{kB}^{\dagger}
\end{pmatrix} \bm h(k-qA(t))
\begin{pmatrix}
c_{kA}\\c_{kB}
\end{pmatrix},
\label{3}
\end{equation}
\begin{equation}
 \bm h(k)=Q_{x}\cos\left({\frac{k}{2}}\right)\boldsymbol{\sigma}_{x}+Q_{y}\sin\left({\frac{k}{2}}\right)\boldsymbol{\sigma}_{y}+Q_{on
 }\boldsymbol{\sigma}_{z},
\end{equation}
where we set $a=\frac{1}{2}$ and
${\bm \sigma}_m\;(m=x,y,z)$
are the Pauli matrices. 
In the SBE approach, we focus on the single-particle density matrix (SPDM)
$\rho_{\alpha\beta,k}^{\rm D}(t)=\langle c_{k\beta}^{\dagger}(t)c_{k\alpha}(t)\rangle$,
where $\langle \cdots \rangle$ is the expectation value
with the grand canonical ensemble and
$c^{\dagger}(t)$ indicates the Heisenberg representation of $c^{\dagger}$. 
The von Neumann equation of SPDM (or simply SBE) is expressed as
\begin{equation}
\partial_{t} \bm \rho_{k}^{\rm D}(t) = -i[ \bm h(k(t)),\bm \rho_{k}^{\rm D}(t)]+\partial_{t}\bm \rho_{k}^{\rm D}|_{\rm relax},
\end{equation}
where $ \bm \rho_{k}^{\rm D}(t)$ is the matrix with elements $\rho_{\alpha\beta,k}^{\rm D}(t)$, $k(t)= k-qA(t)$ and $\hbar$ is set unity. 
The last term represents relaxation and dephasing processes originating from electron-electron interactions, electron-phonon interactions and scattering with impurities.
The microscopic evaluation of $\partial_{t}\bm \rho_{k}^{\rm D}|_{\rm relax}$ is computationally expensive. Instead, in this paper, we set $\partial_{t}\bm \rho_{k}^{\rm D}|_{\rm relax} =  -\frac{\bm {\rho}_{k}^{\rm D}(t)-\bm {\rho}_{\rm{eq},\textit{k}(t)}^{\rm D}}{T_{1}}$ and take account of the relaxation and dephasing processes phenomenologically. Here $\bm {\rho}_{\rm{eq},\textit{k}}^{\rm D}$ represents the SPDM in the equilibrium state. 
In this representation, the operator of the current is expressed as 
\begin{equation}
\hat J(t)=\sum_{k} q \hat \psi_{k}^{\dagger}[\partial_{k}\bm h(k(t))] \hat \psi_{k},
\end{equation}
where $\hat \psi_{k}^{\dagger}=[c_{kA}^{\dagger},c_{kB}^{\dagger}]$. 
Note that the expectation value of the current can be obtained by the SPDM. 
The intensity of HHG is also evaluated from the current $J(t)$
as $I_{HHG}(\omega)=|\omega  J(\omega)|^{2}$,
where $J(\omega)$ is the Fourier component of $J(t)$.
Since the expression of ${\bm h}(k)$ is easily evaluated,
this representation is beneficial for the numerical simulation of
the time evolution of the system excited by the light. 
However, to classify distinct contributions to HHG,
it is more convenient to consider the basis set
that diagonalizes ${\bm h}(k(t))$ as the following representations.

\subsection{Representation II: Dipole gauge expressed with the Houston basis}

We introduce the unitary matrix $\bm  U(k)$ that satisfies $\bm U(k)^{\dagger}\bm h(k)\bm U(k) = \bm \epsilon(k)$, where $\bm \epsilon(k) = \rm{diag}[\epsilon_{c}(\textit{k}),\epsilon_{v}(\textit{k})]$. The subscripts $c$ and $v$ refer to the conduction band  and the valence band, respectively. Considering the time dependent unitary transformation as  ${\hat \psi}_k \rightarrow {\bm U}(k(t)) {\hat \psi}_k$, we obtain the Hamiltonian 
\begin{equation}
  \hat{H}^H(t)=\sum_{k}\hat \psi_{k}^{\prime \dagger}\bm \epsilon(k(t))\hat \psi_{k}^{\prime}-qE(t)\sum_{k}\hat \psi_{k}^{\prime \dagger}\bm d(k(t))\psi_{k}^{\prime},
\end{equation}
where $\hat \psi_{k}^{\prime}=[b_{kc}^{\dagger},b _{kv}^{\dagger}]$.
$\bm d(k) = i\bm U(k)^{\dagger}[\partial_{k}\bm U(k)]$ is the Berry connection, which plays the role of the dipole matrix. 
 In this representation, the von Neumann equation for $\rho_{mn,k}^{\rm H}(t)=\langle b_{kn}^{\dagger}(t)b_{km}(t)\rangle$ becomes
 \begin{equation}
\partial_{t}\bm {\rho}_{k}^{\rm H}(t)= -i[\bm \epsilon(k(t))-qE(t)\bm d(k(t)),\bm \rho_{k}^{\rm H}(t)]. \label{eq:SBE_II}
 \end{equation}
 Actually, this representation is essentially equivalent to the representation III,
which will be shown below.

\subsection{Representation III: Length gauge expressed with the band basis}

Now we express the Hamiltonian (\ref{3}) in the length gauge,
using the band basis $[c_{kc}^{\dagger},c_{kv}^{\dagger}]=[c_{kA}^{\dagger},c_{kB}^{\dagger}]\bm U(k)$.
$\hat H^{L}(t)$ is given as 
\begin{equation} 
\hat H^{\rm{L}}(t)=\sum_{k}
\begin{pmatrix}
c_{kc}^{\dagger}~~c_{kv}^{\dagger}
\end{pmatrix} \bm \epsilon(k)
\begin{pmatrix}
c_{kc}\\c_{kv}
\end{pmatrix}
-E(t) \cdot \hat P,
\end{equation}
where $\hat P$ is the polarization operator.
This operator can be divided into the intra- and interband polariztions
as $\hat P= \hat {P}_{\rm ra} + \hat{P}_{\rm er}$, where
\begin{align} 
\hat{P}_{\rm ra} &= \hat { P}_{\rm ra}^{(\rm{I})}+ \hat {P}_{\rm ra}^{(\rm{II})} \nonumber\\
&= q\sum_{k}\sum_{n} d_{nn}(k)c_{kn}^{\dagger}c_{kn} + q\sum_{k,k^{\prime}}\sum_{n}[i \nabla_{k}\delta(k-k^{\prime})]c_{kn}^{\dagger}c_{k^{\prime}n}, \\
\hat{P}_{\rm er}&=q\sum_{k}\sum_{n \neq m} d_{nm}(k)c_{kn}^{\dagger}c_{km},
\end{align}
where $\bm d(k) = i\bm U(k)^{\dagger}[\partial_{k}\bm U(k)]$.
The current is expressed as the change of the polarization as $\hat J(t) = -i[\hat{P},\hat H^{\rm{L}}(t)]$.
Thus, the intraband current is defined as $\hat J_{\rm ra}(t)=-i[\hat{P}_{\rm ra},\hat H^{\rm{L}}(t)]$ and the interband current is defined as $\hat J_{\rm er}(t)=-i[\hat{P}_{\rm er},\hat H^{\rm{L}}(t)]$. 
The intraband current $ \hat J_{\rm ra}(t) =  \hat J_{\rm ra}^{(\rm{I})}(t) +\hat J_{\rm ra}^{(\rm{II})}(t)$ is given by
\begin{align} 
&\hat J_{\rm ra}^{(\rm{I})}(t) = q\sum_{k}\sum_{n}\partial_{k}\bm \epsilon(k)c_{kn}^{\dagger}c_{kn},\\
&\hat J_{\rm ra}^{(\rm{II})}(t) =-qE(t)\sum_{k}\sum_{n\neq m}\Bigl(\partial_{k}[d_{nm}(k)]-i(d_{nn}(k)-d_{mm}(k))d_{nm}(k)\Bigr)c_{kn}^{\dagger}c_{km}.
\end{align}
We find that
$\hat J_{\rm ra}^{(\rm{I)}}(t)$ consists of the diagonal components of $c^\dagger_n c_m$,
while $\hat J_{\rm ra}^{(\rm{II})}(t)$ consists of the off-diagonal components.
The current $\hat J_{\rm ra}^{(\rm{II})}(t)$ originates from $-i[\hat {P}_{\rm ra},-E(t)\cdot  \hat {P}_{\rm er}]$, and
represents the change of the intraband dipole via interband transition.
The interband current $\hat J_{\rm er}(t)$ is given as 
\begin{equation}
\hat J_{\rm er}(t)=-iq\sum_{k}
\begin{pmatrix}
c_{kc}^{\dagger}&c_{kv}^{\dagger}
\end{pmatrix}
[ {\bm d}(k),\boldsymbol{\epsilon}(k)]
\begin{pmatrix}
c_{kc}\\c_{kv}
\end{pmatrix}
+ qE(t)\sum_{n\neq m}\sum_{k}\Bigl(\partial_{k}[d_{nm}(k)]-i(d_{nn}(k)-d_{mm}(k))d_{nm}(k)\Bigr)c_{kn}^{\dagger}c_{km}.
\end{equation}
The von Neumann equation for $\rho_{mn,k}^{\rm{LB}}(t)=\langle c_{kn}^{\dagger}(t)c_{km}(t)\rangle$ in this representation is 
\begin{equation}
\partial_{t}\bm \rho_{k}^{\rm{LB}}(t)=  -i [\bm h^{\rm{LB}}(k,t),\bm \rho_{k}^{\rm{LB}}(t)]-(E(t)\cdot \nabla_{k})\bm \rho_{k}^{\rm{LB}}(t), \label{eq:SBE_basic}
\end{equation}
where $\bm h^{\rm{LB}}(k,t)=\bm \epsilon(k)-E(t)\bm d(k)$. 
Introducing $\tilde{\bm {\rho}}_{k}^{\rm{LB}}(t)\equiv \bm \rho_{k-qA(t)}^{\rm{LB}}(t)$, we have
\begin{equation}
\partial_{t}\tilde{\bm {\rho}_{k}}^{\rm{LB}}(t)=  -i [\bm h^{\rm{LB}}(k-qA(t),t),\tilde{\bm {\rho}}_{k}^{\rm{LB}}(t)].
\end{equation}
This equation is the same as Eq.\eqref{eq:SBE_II} for SPDM in the representation II.
Since the initial condition of SPDM is also the same
for the representations II and III,
we have $\bm \rho_{k}^{\rm H}(t)=\tilde{\bm {\rho}}_{k}^{\rm{LB}}(t)$.

The SBE in the form of Eq. \eqref{eq:SBE_basic} has often been used up to now,
where the intraband and interband currents are evaluated separately.
However, in this treatment, some terms of the current have been often overlooked~\cite{Wilhelm2021PRB,Murakami2022b},
{\it eg.} the current contribution originating from the change of the intraband dipole
via the interband transition $\hat J_{\rm ra}^{(\rm{II})}(t)$.
In the following, we examine the contributions of different types of currents,
performing the simulation based on the representation I.

\section{Results}
\begin{figure}[H]
\centering
\includegraphics[scale=0.50]{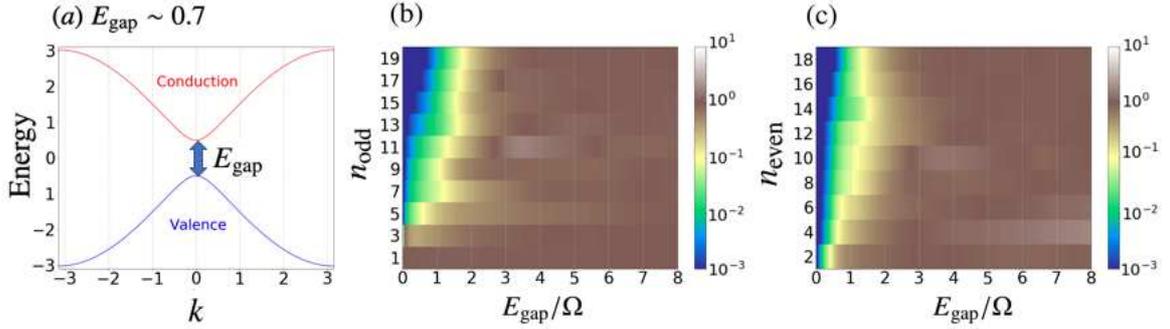}
\caption{(a) Band structure of the Rice-Mele model with $Q_{y}=3$, and $Q_{x}=Q_{on}=0.25$. $E_{\rm gap}$ indicates the minimum band gap.  (b) $\alpha_n$ as a function of $n_{\rm odd}$ and $E_{\rm gap}/\Omega$. (c) $\alpha_n$ as a function of $n_{\rm even}$ and $E_{\rm gap}/\Omega$. For (b) and (c), we set  $Q_{y}=3$ and the chemical potential $\mu=0$ (the half filling), and keep $Q_{x} = Q_{on}$.   The value of $E_{\rm gap}$ reflects the change of the parameters $Q_{x}$ and $Q_{on}$. The parameters of the electric field are  $\Omega$ = 0.2,  $A_{0} = 1$,  $\sigma= 300$, $t_{0}$ = 1200 and $T_{1}$ = 4$\pi$. }
\label{fig1}
\end{figure}

To discuss how important the well-neglected current $\hat J_{\rm ra}^{(\rm{II})}(t)$
is in the HHG analysis,
we examine the time-evolution of the system
after the electric field pulse is introduced.
Here, we consider the electric field pulse in the Gaussian form as
$A(t)=A_{0}\exp({-\frac{(t-t_{0})^{2}}{2\sigma^{2}}})\sin(\Omega(t-t_{0}))$.
To reveal the role of $\hat J_{\rm ra}^{(\rm{II})}(t)$ in HHG,
we introduce $\alpha_n$ as 
\begin{equation}
\alpha_n=\left| \frac{\int_{(n-\delta)\Omega}^{(n +\delta)\Omega}d\omega ~J(\omega)}{\int_{(n -\delta)\Omega}^{(n +\delta)\Omega}d\omega~J_{\rm simp}(\omega)} \right|,
\end{equation}
where $\hat J_{\rm simp}(t)[=\hat J_{\rm ra}^{(\rm{I})}(t)+\hat J_{\rm er}(t)]$
is a part of the full current operator and
$J_{\rm simp}(\omega)$ is the Fourier component of its expectation value.
In practice, we set $\delta=0.5$.
The contribution from $\hat J_{\rm ra}^{(\rm{II})}(t)$ is crucial
when $\alpha_n$ is far away from unity.
First, we focus on the half-filled system ($\mu=0$)
and study the band gap dependence under the condition $Q_x=Q_{on}$.
In the case $Q_y=3$ and $Q_x=Q_{on}=0.25$,
its band structure is shown in Fig.~\ref{fig1}(a),
where the band gap $E_{\rm gap}\sim 0.7$.
In the system, when both $Q_x$ and $Q_{on}$ are small,
the system approaches the system with the inversion symmetry,
and thereby the intensity of odd harmonics is larger than even one. 
In Figs.~\ref{fig1}(b) and \ref{fig1}(c), we show the result for $\alpha_n$ with odd and even $n$, respectively.
We find that $\alpha_n$ becomes much smaller than unity
when $E_{\rm gap}$ is smaller than or comparable to the excitation frequency.
On the other hand, when $E_{\rm gap} $ is large, 
$\alpha_n$ is close to unity.
These results imply that, when the system is half filled and
the band gap is smaller than or comparable to the excitation frequency,
the cancelation between the intraband and interband currents is severe and 
the careful treatment of $\hat J_{\rm ra}(t)$ is needed
to evaluate the HHG spectrum correctly.

\begin{figure}[H]
\centering
\includegraphics[scale=0.54]{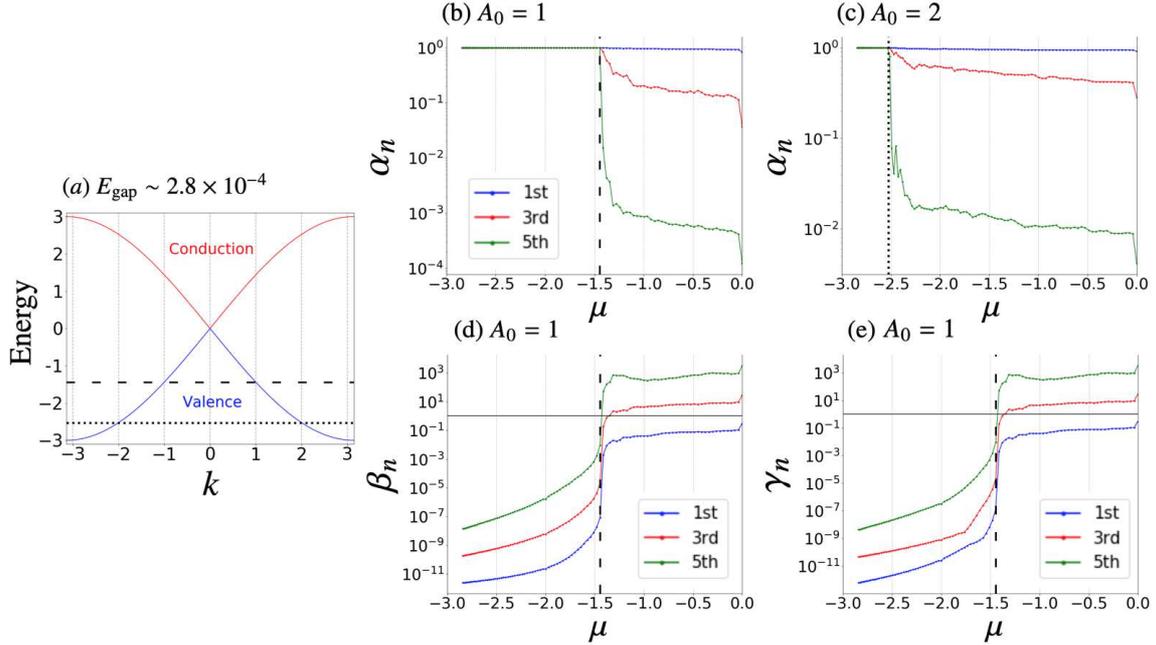}
\caption{(a) Band structure of the model with $Q_{y}=3$ and $Q_{on}=Q_{x}=1\times 10^{-4}$ for which we study the doping dependence of HHG. (b,c) The corresponding $\alpha_n$ as a function of chemical potential.  
(d) The corresponding $\beta_n$ as a function of chemical potential. (e) The corresponding $\gamma_n$ as a function of chemical potential. In all cases, the parameters of the electric field are $\Omega$ = 0.2, $\sigma= 300$, $t_{0}$ = 1200 and $T_{1}$ = 4$\pi$.}
\end{figure}

Next, we discuss the doping dependence of $\hat J_{\rm ra}^{(\rm{II})}(t)$, 
considering the system with $Q_y=3$ and $Q_{on}=Q_{x}=1\times 10^{-4}$,
where a tiny gap appears in the single-particle spectrum, as shown in Fig.~2(a).
In the case, the system is almost symmetric
under the inversion operation. Therefore, we focus on relevant odd harmonics.
Figures~2(b) and 2(c) show $\alpha_n$ as a function of the chemical potential
for odd $n$. 
We find that, as for the first harmonics, $\alpha_1$ is always close to unity,
implying that $\hat J_{ra}^{(\rm II)}(t)$ is irrelevant.
On the other hand, different behavior appears in the higher harmonics ($n=3,5$).
When the system is half filling ($\mu=0$),
$\alpha_n$ is far away from unity, as discussed above.
When the holes are doped in the system,
$\alpha_n$ slightly increases.
When $\mu\simeq \mu_c$,
$\alpha_n\;(n=3,5)$ increases suddenly, and it reaches unity, as shown in Figs.~2(b) and 2(c). 
Here, we introduce $\mu_c$ as $\epsilon_{v}(k=\pm A_{0})$.
These results suggest that the contribution of $\hat J_{\rm ra}^{(\rm{II})}(t)$
is crucial when the system is close to the half filling and
the band gap is small enough to the excitation frequency.

Now we study the distinct contributions for the currents in detail.
To this end, we introduce
\begin{eqnarray}
  \beta_n=\left|\frac{\int_{(n-\delta)\Omega}^{(n+\delta)\Omega}d\omega ~ J_{\rm er}(\omega)}{\int_{(n-\delta)\Omega}^{(n+\delta)\Omega}d\omega~J_{\rm ra}^{(\rm{I})}(\omega)} \right|,&&\;\;\;
  \gamma_n=\left|\frac{\int_{(n-\delta)\Omega}^{(n+\delta)\Omega}d\omega ~ J_{\rm ra}^{(\rm II)}(\omega)}{\int_{(n-\delta)\Omega}^{(n+\delta)\Omega}d\omega~J_{\rm ra}^{(\rm{I})}(\omega)} \right|,
\end{eqnarray}
which allows us to discuss the role of the currents $\hat J_{\rm er}$(t) and $\hat J^{(\rm II)}_{ra}(t)$
in the HHG spectrum.
We show these quantities as a function of the chemical potential
in Figs.~2(d) and 2(e). 
It is found that $\beta_n$ and $\gamma_n$ are qualitatively similar to each other.
Namely, when $\mu \gtrsim \mu_c$, they are almost constant,
while they decrease abruptly around $\mu \simeq \mu_c$.
The results imply that the contribution of $\hat J_{\rm ra}^{(\rm{I})}(t)$ is dominant for $\mu \lesssim \mu_c$, and both $\hat J_{\rm ra}^{(\rm{II})}(t)$ and $\hat J_{\rm er}(t)$ are dominant for $\mu \gtrsim \mu_c$.
This change of the dominant contribution in HHG around $\mu_c(=\epsilon_{v}(k=\pm A_{0}))$ can be understood as follows.
In the present system, one can expect that the interband transition
is strongly suppressed when $\mu \lesssim \mu_c$.
This is because no electrons can reach the Gamma point, where the band gap is minimum, during the pulse.
Note that when an electron is excited by the electric field,
its momentum is shifted by the vector potential.
In addition, $\hat J_{\rm ra}^{(\rm{II})}(t)$ corresponds to the change of the intraband dipole via interband transition and
$\hat J_{\rm er}(t)$ corresponds to the change of the interband dipole.
Therefore, the suppression of the interband transition implies that
these contributions become less important.

\section{Summary}
To summarize, we have studied the effects of the often-neglected current $\hat J_{\rm ra}^{(\rm{II})}(t)$, which originates from the change of the intraband dipole via interband transition, on HHG in the one-dimensional Rice-Mele model. When the system is close to the half filling and the band gap is smaller than or comparable to the excitation frequency, the contribution becomes crucial. 
Our results suggest the importance of the full evaluation of the currents when one studies HHG in small gap systems such as graphene, Weyl semimetals and metallic carbon nanotubes. \\

\section*{Acknowledgment}
We would like to acknowledge fruitful discussions with Michael Sch\"uler. This work is supported by Grant- in-Aid for Scientific Research from JSPS, KAKENHI Grant Nos. JP20K14412, JP21H05017 (Y. M.),
JP17K05536, JP19H05821, JP21H01025, JP22K03525 (A.K.),
JST CREST Grant No. JPMJCR1901 (Y. M.).

\end{document}